\documentclass[letterpaper,nofootinbib,prd,amsmath]{revtex4}

\usepackage[dvips]{graphics,color}
\usepackage{epsfig}
\usepackage{hyperref}
\newcommand\dperp{{\partial}_\perp}
\newcommand\LL{{I{\hskip-3pt}L}}

\begin{document}

\title{Conformal invariance and the
conformal--traceless decomposition\\ of the gravitational field}
\author{J.~David Brown}
\affiliation{Department of Physics, North Carolina State University,
Raleigh, NC 27695 USA}

\begin{abstract}
Einstein's  theory of general relativity is written in terms of the
variables obtained from a conformal--traceless decomposition of the spatial metric and extrinsic 
curvature. The determinant of the conformal metric is not restricted, so the action functional 
and equations of motion  are invariant under  conformal transformations. With this approach the
conformal--traceless variables remain free of density weights. The conformal invariance of the 
equations of motion can be broken by imposing an evolution equation for the determinant 
of the conformal metric $g$. Two conditions are considered, one in which $g$ is constant 
in time and one in which $g$ is constant along the unit normal to the spacelike 
hypersurfaces. This approach is used to write the Baumgarte--Shapiro--Shibata--Nakamura 
system of evolution equations in conformally invariant form. The presentation includes a 
discussion of the
conformal thin sandwich construction of gravitational initial data, and the conformal flatness 
condition as an approximation to the evolution equations. 
\end{abstract}
\maketitle

\section{Introduction}
The conformal--traceless decomposition of the gravitational field was originally introduced 
by Lichnerowicz \cite{Lichnerowicz} and York \cite{York:1972sj, York:1979} in their work 
on the initial value problem. Since that time the same decomposition has appeared in various 
guises in mathematical and numerical relativity. The BSSN \cite{Shibata:1995we,Baumgarte:1998te} 
(Baumgarte--Shapiro--Shibata--Nakamura) system of evolution equations for general 
relativity is currently in widespread use in the numerical relativity community. The BSSN 
equations are based on a conformal--traceless splitting.
The recently discovered conformal thin 
sandwich construction \cite{York:1998hy,Pfeiffer:2002iy}, like the earlier techniques for 
solving the initial data problem, uses the conformal--traceless decomposition. Unlike its 
predecessors, the conformal thin sandwich scheme mixes a subset of the evolution equations 
with the constraints to facilitate the analysis.  The conformal flatness condition (CFC) is 
an approximation to the Einstein equations that uses the conformal--traceless decomposition. It has been 
applied to numerical simulations of binary neutron star systems \cite{Wilson:1996ty} and supernovae 
\cite{Dimmelmeier:2002bk}. 

The conformal--traceless (CT) decomposition is 
\begin{equation}\label{ctdecomposition}
  h_{ab} = \varphi^4 g_{ab} \ ,\qquad K_{ab} = \varphi^{-2} A_{ab} + \frac{1}{3}\varphi^4 g_{ab} \tau \ ,
\end{equation}
where $h_{ab}$ is the physical spatial metric and $K_{ab}$ is the physical extrinsic curvature. The definitions 
(\ref{ctdecomposition}) are redundant, in the sense that multiple sets of fields $g_{ab}$, $\varphi$, $A_{ab}$, $\tau$ 
yield the same physical metric and extrinsic curvature. To be precise, these definitions are invariant under the 
conformal transformation\footnote{The arrow notation means the following: Replace the fields $g_{ab}$, $\varphi$, 
$A_{ab}$, and $\tau$ with their ``barred'' counterparts, which differ  from the unbarred fields by certain factors 
of $\xi$.  In Eq.~(\ref{ctdecomposition}) the factors of 
$\xi$ completely cancel in each term so that $h_{ab}$ and $K_{ab}$ are independent of $\xi$. Thus, 
$h_{ab}$ and $K_{ab}$ are conformally invariant.}
\begin{subequations}\label{conformaltransformations}
\begin{eqnarray}
  g_{ab} & \longrightarrow & {\bar g}_{ab}  =  \xi^4 g_{ab} \ ,\\
  \varphi & \longrightarrow & {\bar \varphi}  =  \xi^{-1} \varphi \ ,\\
  A_{ab} & \longrightarrow & {\bar A}_{ab} = \xi^{-2} A_{ab} \ ,\\
  \tau & \longrightarrow & {\bar \tau} = \tau \ ,
\end{eqnarray}
\end{subequations}
for any field $\xi$. 
In many recent applications involving the conformal--traceless splitting the conformal invariance 
(\ref{conformaltransformations}) is broken, and the redundancy in the 
CT variables removed, by  the 
condition that $g_{ab}$ should have unit determinant. There are various insights to be gained by leaving the determinant 
of $g_{ab}$ unspecified at this point in the analysis. Thus, {\it I will not impose} $g \equiv \det(g_{ab}) = 1$. 

In most of the original works on the initial value problem and on gravitational degrees of freedom 
\cite{York:1971hw,York:1972sj} the determinant 
$g$ is not restricted. Rather, the field $g_{ab}$ is described as representing a conformal equivalence class of 
metrics that defines a conformal geometry. The approach taken here is equivalent. However, instead of 
describing the results in terms of conformal geometry, I treat $g_{ab}$ as an ordinary spatial metric (symmetric 
type $0\choose 2$ tensor) and emphasize the conformal invariance (\ref{conformaltransformations}) of 
the action and equations of motion. 

The physical metric and extrinsic curvature are also invariant under the  ``trace transformation''  
defined by $\tau \longrightarrow {\bar \tau} = \tau + \zeta$
and $A_{ab}  \longrightarrow  {\bar A}_{ab} =  A_{ab} - \zeta \varphi^6 g_{ab} /3$, with $g_{ab}$ and $\varphi$ unchanged,
for any field $\zeta$.
This redundancy in the CT variables is usually removed by the condition that $A_{ab}$ should 
have zero trace. It turns out that there is no particular advantage to be gained by leaving the trace of $A_{ab}$ 
unspecified. Thus, 
{\it I will impose} $A \equiv A_{ab}g^{ab} = 0$. Then the CT variable $\tau$ is the trace of the extrinsic  
curvature, $\tau = K_{ab}h^{ab}$. 

At first sight it might appear surprising that I have chosen to keep the conformal invariance but eliminate the 
trace invariance. The reason is  that the trace transformation is essentially trivial. It just corresponds
to a change in the splitting of the trace of $K_{ab}$ into the two terms $\tau$ and $A\varphi^{-6}$. 
With the choice  $A=0$, the trace of 
the extrinsic curvature is placed entirely in $\tau$. In a similar fashion, the conformal transformation  can be viewed as 
a change in the splitting of the determinant of $h_{ab}$ into the two factors $\varphi^{12}$ and  $g$. 
However, this is a nontrivial change precisely because the field $\xi$ can carry a nonzero density weight. Throughout 
this paper I 
define the ``unbarred'' CT variables that appear in the definitions (\ref{ctdecomposition}) as having no density weight. 
Thus, $g_{ab}$ and $A_{ab}$ are spatial tensors of 
type $0\choose 2$, and $\varphi$ and $\tau$ are spatial scalars. If $\xi$ carries a density weight, then 
the conformally transformed variables, the ``barred'' variables in Eqs.~(\ref{conformaltransformations}), 
acquire density weights. 

By not specifying $g=1$ from the outset, the equations of motion we obtain are conformally invariant. We can then 
consider breaking the conformal invariance by fixing $g$. In principle $g$ can be chosen as any $t$--dependent 
scalar density of weight $2$. In practice I expect that the most natural way to break conformal invariance is to
choose $g$ arbitrarily on the initial time slice and  then evolve $g$ according to some simple prescription. Two natural 
prescriptions are considered, one in which $g$ remains fixed along the time flow vector field (the ``Lagrangian condition'')
and one in which $g$ remains fixed along the normal to the spacelike hypersurfaces (the ``Eulerian condition''). 
The equations of motion differ between these two cases in the way that the shift terms appear. These differences 
can be interpreted as changes in the density weights of the CT variables. 

In this paper I carefully examine the full Einstein theory, constraints and  evolution equations, in terms of the
conformal--traceless variables. One of my motivations for this analysis is to help clarify various subtle issues 
that arise along the way. Most of the subtleties concern density weights. 
Consider what happens when $g=1$ is fixed from the outset. Then the conformal metric and conformal factor must 
carry nonzero density weights, because otherwise $g$ would not remain equal to $1$ under a change of spatial coordinates. 
This leads to some unusual and potentially confusing results. For example, in the initial value 
problem the Hamiltonian constraint is written in terms of the conformal Laplacian operator acting on the 
conformal factor. Yet, the conformal Laplacian acts  as if neither the 
conformal metric nor the conformal factor carries any density weight.  It is clear that 
the experts on the initial value problem already recognize and understand this subtle point. 
One of my goals for this paper is to present a framework for the conformal--traceless decomposition in 
which these subtle issues are largely avoided.

A second motivation for this analysis is to help pave the way for the development of numerical techniques 
in general relativity based on the ``variational integrator'' approach to numerical modeling. (For an overview, 
see Ref.~\cite{VIoverview}.) In this approach, 
finite difference equations are derived by extremizing the action 
functional in discrete form. It turns out that for a constrained system like general relativity, the  equations 
obtained from a discrete variational principle will mix the constraints and the evolution equations at each timestep. 
If the fundamental variables are the physical metric and extrinsic curvature, then one is faced with
the problem of solving a discrete version of the thin sandwich equations \cite{ThinSandwich}. 
By now it is well known that the 
thin sandwich equations are ill--posed, and do not admit a solution in the generic case. On the other hand, the conformal 
thin sandwich equations are expressed in terms of ``nice'' elliptic operators and one expects the system to be 
generically well--posed. In order to apply the variational integrator construction to general 
relativity, we must first express the action in terms of the conformal--traceless variables. 
 
Throughout this paper I  use the lapse anti--density $\alpha$ as  the undetermined 
multiplier for the Hamiltonian constraint. The lapse anti--density carries density weight $-1$ and the 
Hamiltonian constraint carries density weight $2$. York and collaborators 
\cite{Anderson:1998we, York:1998hy, Pfeiffer:2002iy} have pointed 
out a number of reasons why this choice is preferred over the traditional scalar lapse function. 
It has also been shown \cite{Sarbach:2002bt} that the BSSN  equations are equivalent to a strongly hyperbolic system 
with physical characteristic speeds when the lapse has density weight $-1$.

In Sec.~II I begin by writing down the action and equations of motion in Hamiltonian form. 
In Sec.~III the action is expressed in terms of the CT variables (\ref{ctdecomposition}). 
Because the action is conformally invariant, it can also be expressed in terms of a set of conformally invariant variables 
that includes the density weight $-2/3$ metric ${\tilde g}_{ab} = g^{-1/3} g_{ab}$. The equations of 
motion for the CT variables are derived in Sec.~IV. There, it is pointed out that the conformal invariance can be broken with 
either the Lagrangian or Eulerian condition on the determinant of $g_{ab}$. This leads to two  
sets of evolution equations that differ in the way that the shift terms enter. In Sec.~V I examine the BSSN 
system. It is written first in conformally invariant form, then with the Eulerian and Lagrangian conditions 
applied. The conformal thin sandwich equations are derived in Sec.~VI. They follow directly from the Hamiltonian and 
momentum constraints and the evolution equations of Sec.~IV. The CFC approximation is 
discussed in Sec.~VII. There it is pointed out that the CFC equations are identical to  the conformal thin 
sandwich equations with the restrictions that the conformal metric $g_{ab}$ is flat and the time slicing 
is maximal, $\tau = 0$. The main results are summarized briefly in Sec.~VIII. 

\section{ADM action and equations of motion}
The Arnowitt--Deser--Misner (ADM) gravitational action is \cite{Arnowitt:1962hi}
\begin{equation}\label{eqn:ADMaction}
   S^{(g)}[h_{ab},P^{ab},\alpha,\beta^a] = \int dt\,d^3x \Bigl[ P^{ab} \dot{h}_{ab} - \alpha {\cal H}^{(g)} - 
   \beta^a {\cal M}^{(g)}_a \Bigr] \ ,
\end{equation}
which is a functional of the physical 3--metric $h_{ab}$, its conjugate $P^{ab}$, the lapse anti--density $\alpha$, and the 
shift vector $\beta^a$. The dot denotes a time derivative, ${\dot h}_{ab} \equiv \partial h_{ab}/\partial t$. The gravitational 
momentum is related to the extrinsic curvature by 
\begin{equation} \label{eqn:Pdefinition}
  P^{ab} = \frac{1}{2} \sqrt{h} ( h^{ab}h^{cd} - h^{ac} h^{bd}) K_{cd} \ .
\end{equation}
The gravitational Hamiltonian and momentum densities are defined by 
\begin{subequations}\label{eqn:ADMdensities}
\begin{eqnarray} 
   {\cal H}^{(g)} & \equiv &  2 P^{ab} P_{ab} - P^2 - h{\cal R}/2 \ ,\label{ADMHamiltoniandensity} \\
   {\cal M}^{(g)}_a & \equiv & - 2\nabla_b P^b_a \ ,\label{ADMmomentumdensity}
\end{eqnarray}
\end{subequations}
where $P = P^{ab} h_{ab}$ is the trace of the gravitational momentum and $h$ is the determinant of $h_{ab}$. Also, 
$\nabla_a$ and ${\cal R} \equiv h^{ab} {\cal R}_{ab}$ are the covariant derivative and scalar curvature for the 
spatial metric $h_{ab}$. If matter fields 
are present the complete action $S = S^{(g)} + S^{(m)}$  includes a functional $S^{(m)}$ of the matter fields 
in addition to the gravitational action. 
I will assume that the matter is ``minimally coupled'' to gravity, so the matter action does not depend on derivatives of the 
spatial metric or on $P^{ab}$. Throughout this
paper I use units in which $8\pi G = 1$, where $G$ is Newton's constant. 

Variations of the ADM action (\ref{eqn:ADMaction}) plus matter action $S^{(m)}$ with respect to the lapse anti--density and 
shift vector yield the constraints
\begin{equation} \label{eqn:constraints}
   {\cal H}^{(g)} = - h\rho  \ ,\qquad {\cal M}^{(g)}_a = \sqrt{h} j_a \ ,
\end{equation}
where $\rho$ and $j_a$ are the energy and momentum densities for the matter fields. They are defined, along with the 
spatial stress tensor $s^{ab}$, in terms of the functional derivatives of the matter action by 
\begin{equation}\label{eqn:matterdefinitions}
   \rho \equiv -\frac{1}{h} \frac{\delta S^{(m)}}{\delta\alpha} \ ,\qquad j_a \equiv \frac{1}{\sqrt{h}} 
   \frac{\delta S^{(m)}}{\delta \beta^a} \ ,
   \qquad s^{ab} \equiv \frac{2}{\alpha h} \frac{\delta S^{(m)}}{\delta h_{ab}} - \frac{h^{ab}}{h} 
   \frac{\delta S^{(m)}}{\delta \alpha}  \ .
\end{equation}
Variations of the action with respect to $h_{ab}$ and $P^{ab}$ yield the well--known ADM evolution equations 
\begin{subequations} \label{eqn:ADMequations}
\begin{eqnarray}
  \dperp h_{ab} & = & \alpha \Bigl( 4 P_{ab} - 2 P h_{ab} \Bigr) \ ,\label{eqn:ADMhdot} \\
  \dperp P^{ab} & = & \alpha \Bigl( -4 P^{ac}P_c^b + 2 P P^{ab} + \frac{1}{2} h{\cal R} h^{ab} - \frac{1}{2} h {\cal R}^{ab} 
       \Bigr)
       + \frac{1}{2} h \nabla^a \nabla^b\alpha - \frac{1}{2} h h^{ab} \nabla^2\alpha 
       + \frac{1}{2} \alpha h (s^{ab} - \rho h^{ab}) \ .
   \label{eqn:ADMPdot}
\end{eqnarray}
\end{subequations}
The derivative operator on the left--hand sides of these equations is defined by 
\begin{equation} \label{eqn:dperp}
   \dperp \equiv \partial/\partial t - {\cal L}_\beta 
\end{equation}
where ${\cal L}_\beta$ is a Lie derivative along the shift vector field $\beta^a$. Thus, $(\alpha\sqrt{h})^{-1}\dperp$ 
is the derivative with respect to proper time along the unit normal to the spacelike hypersurfaces.
Specifically, the Lie derivatives are defined by 
\begin{subequations} \label{eqn:Lie_handP}
\begin{eqnarray}
  {\cal L}_\beta h_{ab} & \equiv & \beta^c \partial_c h_{ab} + h_{ac} \partial_{b}\beta^c + h_{cb} \partial_a \beta^c 
    = 2\nabla_{(a} \beta_{b)} \ ,\label{eqn:Lie_h} \\
    {\cal L}_\beta P^{ab} & \equiv & \partial_c(\beta^c P^{ab}) - P^{ac}\partial_c\beta^b - P^{cb}\partial_c\beta^a 
    = \nabla_c(\beta^c P^{ab}) - 2 P^{c(a} \nabla_c\beta^{b)} \ ,\label{eqn:Lie_P} 
\end{eqnarray}
\end{subequations}
where $\partial_a = \partial/\partial x^a$ and $(a \cdots b)$ denotes symmetrization on the indices $a$ and $b$. 
Note that $h_{ab}$ is a symmetric type ${0\choose 2}$ 
tensor on the spatial manifold, and $P^{ab}$ is a symmetric type ${2\choose 0}$ tensor density of weight $1$ on the 
spatial manifold. (The terminology of density weights 
is fixed by noting that $\sqrt{h}$ is a scalar density of weight $1$.)

The time derivatives in the action (\ref{eqn:ADMaction}) can be written in terms of the normal derivative $\dperp$, with 
the result 
\begin{equation} \label{eqn:ADMaction2}
  S^{(g)}[h_{ab},P^{ab},\alpha,\beta^a] = \int dt\,d^3x \Bigl[ P^{ab} \dperp {h}_{ab} - \alpha {\cal H}^{(g)} \Bigr] \ ,
\end{equation}
The difference between the actions (\ref{eqn:ADMaction}) and (\ref{eqn:ADMaction2}) is a total derivative, 
$2\int dt\,d^3x \nabla_a ( P^{ab}\beta_b)$. This term vanishes for compact spatial manifolds, and integrates to 
a boundary term for non--compact spatial manifolds. The presence of boundary terms in the action affect the admissible 
boundary conditions to be imposed in the variational principle, but they do not affect the resulting equations of motion. 
I will ignore such boundary terms in the present work. 

\section{Action in conformal--traceless variables}
With the  physical spatial metric and extrinsic curvature written as in Eq.~(\ref{ctdecomposition}), the 
gravitational momentum (\ref{eqn:Pdefinition}) becomes 
\begin{equation} \label{ctdecomposition_P}
  P^{ab} = -\frac{1}{2}\sqrt{g} \varphi^{-4} A^{ab} + \frac{1}{3} \sqrt{g}\varphi^2 \tau g^{ab} \ .
\end{equation}
Recall that $A_{ab}$ is trace free, $A = 0$. Also note that 
indices on $A_{ab}$ are raised with the inverse conformal metric $g^{ab}$. We can express the ADM 
action  in terms of the conformal--traceless (CT) variables by substituting the expressions for $h_{ab}$ and $P^{ab}$ 
into Eq.~(\ref{eqn:ADMaction2}). The result is
\begin{equation} \label{eqn:confaction1}
   S^{(g)}[g_{ab},A_{ab},\varphi,\tau,\alpha,\beta^a] = \int dt\,d^3x \left[ -\frac{1}{2} g^{5/6}  A^{ab}
     \dperp \left(g^{-1/3} g_{ab}\right) + \frac{2}{3}  \tau  \dperp\left( g^{1/2}\varphi^6\right) 
    - \alpha {\cal H}^{(g)} \right] \ ,
\end{equation}
where the gravitational contribution to the Hamiltonian constraint is 
\begin{equation}\label{eqn:conformalHdensity}
  {\cal H}^{(g)} \equiv \frac{1}{2} g  A^{ab}A_{ab}  - \frac{1}{3} g \varphi^{12} \tau^2 
   - \frac{1}{2} g \varphi^8 R + 4 g  \varphi^7 D^2\varphi \ .
\end{equation}
Here,  $R$ is the scalar curvature of $g_{ab}$, and $D_a$ is the 
covariant derivative compatible with $g_{ab}$. Note that the ``velocity'' term  $\dperp  \left(g^{-1/3} g_{ab}\right)$ 
in the action is trace free. 
It can be written as $\dperp  \left(g^{-1/3} g_{ab}\right) = g^{-1/3} \left( \dperp g_{ab} \right)^{\rm TF}$, 
where TF stands for the trace--free part of the expression enclosed in parentheses. 

Observe that the traceless property of  $A_{ab}$ must be preserved when the action is varied.
This condition can be enforced with a Lagrange multiplier. The result is equivalent to  simply
demanding that the functional derivative with respect to $A_{ab}$ must be traceless. In general, if
we have a functional $S[T_{ab}]$ of a traceless tensor $T_{ab}$ and $\delta S = \int d^4x\, E^{ab}\delta T_{ab}$,
then extremization of $S[T_{ab}]$
yields $\left( E^{ab} \right)^{\rm TF} = 0$. This result, that the functional derivative
with respect to a traceless tensor is trace--free, should not be too surprising.
For example, we routinely vary
action functionals with respect to symmetric tensors. The symmetry property can be enforced with a Lagrange
multiplier, but the result is equivalent to simply demanding that the functional derivative should be symmetric.
 
A key piece of the calculation for the action (\ref{eqn:confaction1}) is the expression for the 
scalar curvature of the physical metric $h_{ab}$ in terms of the conformal metric $g_{ab}$ and conformal 
factor $\varphi$:
\begin{equation}\label{Rscalar}
  {\cal R} = \varphi^{-4} R - 8\varphi^{-5} D^2 \varphi \ .
\end{equation}
This result can be derived 
in a straightforward way by inserting the decomposition $h_{ab} = \varphi^4 g_{ab}$ into the definition 
for the Ricci scalar, 
${\cal R} = 2 g^{ab}\left( \partial_{[c} \Gamma^c_{b]a} +  \Gamma^c_{d[c}  \Gamma^d_{b]a}\right)$, where $\Gamma^a_{bc} 
= h^{ad} \left( \partial_b h_{dc} + \partial_c h_{bd} - \partial_d h_{bc} \right)/2$ are the Christoffel 
symbols. 

The gravitational field contributions to the momentum constraint are ``hidden'' in the $\dperp$ terms in the action, since 
these terms include Lie derivatives with 
respect to the shift vector $\beta^a$. The Lie derivatives are defined by the tensor character of the 
variables on which they act. For example, ${\cal L}_\beta g_{ab}$ and ${\cal L}_\beta A^{ab}$ are given by 
expressions (\ref{eqn:Lie_handP}) with $h_{ab}$ replaced by $g_{ab}$, $P^{ab}$ replaced by $A^{ab}$, and $\nabla_a$ 
replaced by $D_a$. (Indices on  $\beta^a$ and $D_a$, like $A_{ab}$, are raised and lowered with the conformal 
metric $g_{ab}$ and its inverse $g^{ab}$.) In the calculations here and below, it is convenient to 
use the formula
\begin{equation}\label{Liederivative}
  {\cal L}_\beta T = g^{w/2} {\cal L}_\beta\left( T g^{-w/2} \right) + w T D_c\beta^c 
\end{equation}
for the Lie derivative, where $T$ is a tensor density of weight $w$. (The indices have been suppressed on $T$.) 
Then the shift terms in the action (\ref{eqn:confaction1}) are 
\begin{equation}
  \left.S^{(g)}\right|_{\rm{shift\ terms}} 
  = \int dt\,d^3x \sqrt{g} \Bigl[  A^{ab} \left( D_a\beta_b\right)^{\rm TF}  
    - \frac{2}{3}  \tau  D_a\left( \beta^a \varphi^6 \right) 
    \Bigr] \ ,
\end{equation}
and the gravitational contribution to the momentum constraint is
\begin{equation}\label{conformalMdensity}
  {\cal M}^{(g)}_a \equiv -\frac{\delta S^{(g)}}{\delta \beta^a} = \sqrt{g} D_b  A^b_a 
  - \frac{2}{3} \sqrt{g} \varphi^6 D_a  \tau  \ .
\end{equation}
The full Hamiltonian and momentum constraints are $0 = {\cal H}^{(g)} + g \varphi^{12} \rho$ and 
$0 = {\cal M}^{(g)}_a -\sqrt{g} \varphi^6 j_a$, respectively. They are invariant under the conformal transformation
(\ref{conformaltransformations}). 

The action as written in Eq.~(\ref{eqn:confaction1}) is a functional of the symmetric type $0\choose 2$ tensor 
$g_{ab}$, the symmetric type $0\choose 2$ tensor $A_{ab}$, the scalars $\varphi$ and $\tau$, and the 
lapse anti--density and shift vector. Because the CT variables are redundant the equations of motion 
obtained by varying the action 
are not independent. To be precise, the action is conformally invariant so its variation vanishes when the CT fields 
are varied by an infinitesimal  conformal transformation (\ref{conformaltransformations}).  
This leads to the relation 
\begin{equation}\label{eomrelation}
  0 = 4 g_{ab} \frac{\delta S^{(g)}}{\delta g_{ab}} - \varphi \frac{\delta S^{(g)}}{\delta\varphi} 
     - 2 A_{ab} \frac{\delta S^{(g)}}{\delta A_{ab}} 
\end{equation}
for  all field configurations, not just those that satisfy the classical equations of motion. 

The conformal invariance of the theory defined by the action (\ref{eqn:confaction1}) can be displayed in 
an elegant form by treating it as a constrained Hamiltonian system \cite{Henneaux:1992ig}. In that case the action becomes 
\begin{equation}
  S^{(g)} = \int dt\, d^3x \,\left[ {\cal P}_{\rm metric}^{ab}\, \dperp {\cal Q}_{ab}^{\rm metric} + 
    {\cal P}_{\rm phi}\, \dperp {\cal Q}^{\rm phi} + {\cal P}_{\rm rootg}\, \dperp {\cal Q}^{\rm rootg} 
    - \alpha {\cal H}^{(g)} 
    - \epsilon\, {\cal C} \right]  \ ,
\end{equation}
where ${\cal P}_{\rm metric}^{ab}$, ${\cal P}_{\rm phi}$, and ${\cal P}_{\rm rootg}$ are the momenta conjugate to 
the canonical coordinates ${\cal Q}_{ab}^{\rm metric} = g^{-1/3} g_{ab}$, ${\cal Q}^{\rm phi} = \varphi^6$, and 
${\cal Q}^{\rm rootg} = \sqrt{g}$, respectively. These variables are restricted by the first 
class constraint ${\cal C} \equiv {\cal Q}^{\rm phi}{\cal P}_{\rm phi} - {\cal Q}^{\rm rootg} {\cal P}_{\rm rootg}$. 
The constraint ${\cal C}$ appears in the action with an undetermined multiplier $\epsilon$. The momenta 
are related to the original CT variables by ${\cal Q}^{\rm phi} {\cal P}_{\rm phi} = {\cal Q}^{\rm rootg} {\cal P}_{\rm rootg} 
= (2/3)\sqrt{g} \varphi^6 \tau$ and ${\cal P}^{ab}_{\rm metric} = -(1/2) g^{5/6} A^{ab}$. These relations can be used to 
rewrite the Hamiltonian constraint (\ref{eqn:conformalHdensity}) in terms of the canonical variables. 
The ``smeared'' constraint $\int d^3x \, \epsilon\, {\cal C}$ generates an infinitesimal conformal transformation 
(\ref{conformaltransformations}) 
through the Poisson brackets with $\xi = 1 - \epsilon/6$. 

Let us return to the action as expressed in Eq.~(\ref{eqn:confaction1}). The equations of motion obtained from 
this action are redundant, as shown by Eq.~(\ref{eomrelation}), because the CT variables are not unique. Said another way, 
the action (\ref{eqn:confaction1}) does not depend on $g_{ab}$, $\varphi$, $A_{ab}$, and $\tau$ separately but 
only on the combinations 
\begin{subequations}\label{canonicalCTvariables}
\begin{eqnarray}
  {\tilde g}_{ab} & \equiv & g^{-1/3} g_{ab} = h^{-1/3} h_{ab}\ ,\\
  {\tilde \varphi} & \equiv & g^{1/12} \varphi  = h^{1/12}\ ,\\
  {\tilde A}^{ab} & \equiv &  g^{5/6}  A^{ab} = -2 h^{1/3} \left( P^{ab} - P h^{ab}/3 \right)\ ,\\
  {\tilde \tau} & \equiv &  \tau  =  P/\sqrt{h} \ .
\end{eqnarray}
\end{subequations}
These quantities are invariant under the conformal  transformation 
(\ref{conformaltransformations}). I will refer to ${\tilde g}_{ab}$,
${\tilde\varphi}^6$, ${\tilde A}^{ab}$ and ${\tilde\tau}$ as the ``invariant CT variables.''  
Note that ${\tilde g}_{ab}$ is the unit determinant metric. It is a type 
$0\choose 2$ tensor density with weight $-2/3$. 
The variable ${\tilde A}^{ab}$ is a traceless type $2\choose 0$ tensor density with weight 
$5/3$. It is proportional to the 
trace--free part of the extrinsic curvature. The variable  ${\tilde\varphi}$ is a scalar  
density with weight $1/6$, and ${\tilde \tau}$ is the trace of the extrinsic curvature with 
no density weight. Note that indices on ${\tilde A}^{ab}$ and other ``tilde'' quantities are raised and lowered 
with ${\tilde g}_{ab}$ and its inverse ${\tilde g}^{ab}$. 

The invariant CT variables (the variables with tilde's) are combinations of the CT variables 
(the variables without tilde's) that are invariant under the 
conformal transformation (\ref{conformaltransformations}). But they can also be viewed as 
the CT variables  transformed by Eqs.~(\ref{conformaltransformations}) 
with $\xi = g^{-1/12}$. This implies that the action, which is conformally invariant, 
has the same expression in terms of the invariant CT variables as it does in terms of the original CT variables. 
Thus, we can rewrite Eqs.~(\ref{eqn:confaction1}), (\ref{eqn:conformalHdensity}), and (\ref{conformalMdensity}) 
by placing tilde's
on each of the CT variables: 
\begin{equation} \label{eqn:confactiontilde}
   S^{(g)}[{\tilde g}_{ab},{\tilde A}_{ab},{\tilde\varphi},{\tilde\tau},\alpha,\beta^a] 
   = \int dt\,d^3x \left[ -\frac{1}{2} {\tilde A}^{ab}
     \dperp {\tilde g}_{ab} + \frac{2}{3} {\tilde\tau}  \dperp{\tilde\varphi}^6 
    - \alpha {\cal H}^{(g)} \right] \ ,
\end{equation}
where 
\begin{subequations}
\begin{eqnarray}
  {\cal H}^{(g)} & = & \frac{1}{2}  {\tilde A}^{ab}{\tilde A}_{ab}  - \frac{1}{3}{\tilde \varphi}^{12} {\tilde\tau}^2 
   - \frac{1}{2}{\tilde \varphi}^8 {\tilde R} + 4  {\tilde\varphi}^7 {\tilde D}^2{\tilde\varphi} 
   \ ,\label{eqn:conformalHdensitytilde}\\
  {\cal M}^{(g)}_a & = &  {\tilde D}_b  {\tilde A}^b_a 
  - \frac{2}{3}{\tilde \varphi}^6 {\tilde D}_a  {\tilde\tau}  \ .\label{conformalMdensitytilde}
\end{eqnarray}
\end{subequations}
Note that the determinant of ${\tilde g}_{ab}$ has been set to one.  The full Hamiltonian 
and momentum constraints are ${\cal H} = {\cal H}^{(g)} + {\tilde\varphi}^{12}\rho$ and 
${\cal M}_a = {\cal M}^{(g)}_a - {\tilde\varphi}^6 j_a$, respectively. 

In the Hamiltonian density ${\cal H}^{(g)}$, the term 
$\tilde R$ is constructed from ${\tilde g}_{ab}$ according to the usual formula
${\tilde R} = 2{\tilde g}^{ab}\left( \partial_{[c} {\tilde\Gamma}^c_{a]b} 
+ {\tilde\Gamma}^d_{a[b} {\tilde\Gamma}^c_{c]d} \right)$ 
where 
${\tilde\Gamma}^a_{bc} = {\tilde g}^{ad} \left( \partial_b {\tilde g}_{dc} 
+ \partial_c {\tilde g}_{bd} - \partial_d {\tilde g}_{bc} \right)/2$. 
Likewise, the Laplacian operator acting on $\tilde\varphi$ is defined by 
${\tilde D}^2{\tilde\varphi} = {\tilde g}^{ab} \left( \partial_a \partial_b {\tilde\varphi} 
- {\tilde\Gamma}^c_{ab}\partial_c {\tilde\varphi}\right)$. 
Because the metric ${\tilde g}_{ab}$ itself carries 
a density weight it transforms under a change of spatial coordinates with an extra factor of $J^{-2/3}$ as compared 
with a weight--zero metric. Here,
$J$ is the Jacobian of the transformation. As a consequence of the density weighting on ${\tilde g}_{ab}$, 
the terms ${\tilde R}$ and ${\tilde D}^2{\tilde\varphi}$ are not 
separately scalars under spatial coordinate 
transformations. However, the combination 
${\tilde\varphi}^{-4}{\tilde R} - 8{\tilde\varphi}^{-5} {\tilde D}^2{\tilde\varphi}$, which equals the physical 
scalar curvature ${\cal R}$, is a scalar. Then together the terms 
$-{\tilde\varphi}^{8}{\tilde R}/2  + 4 {\tilde\varphi}^{7}{\tilde D}^2{\tilde\varphi}$ 
that appear in the Hamiltonian density  (\ref{eqn:conformalHdensitytilde}) 
transform as a scalar density of weight $2$. Recall that the lapse anti--density $\alpha$ carries weight $-1$, so 
$\alpha {\cal H}^{(g)}$ is a weight $1$ density as it should be. 

Also observe that the Laplacian ${\tilde D}^2$ acts on 
$\tilde\varphi$ in Eq.~(\ref{eqn:conformalHdensitytilde}) as if $\tilde\varphi$ were a scalar rather than a 
scalar density of weight $1/6$. This is because the difference 
between ${\tilde D}_a$ acting on a scalar and  acting on 
a scalar density is a term proportional to $\partial_a {\tilde g}$, which 
vanishes because ${\tilde g}_{ab}$ has unit determinant. Likewise, in the momentum density 
(\ref{conformalMdensitytilde}), the covariant derivative ${\tilde D}_b$  acts on 
${\tilde A}^{b}_a$ as if ${\tilde A}^{b}_a$ were a type $1\choose 1$ tensor with no density weight. 

\section{Equations of motion for the conformal--traceless variables}

Let us derive the dynamical equations of motion by varying the action (\ref{eqn:confactiontilde}) 
with respect to the invariant CT variables. Note that 
the variations  $\delta {\tilde A}^{ab}$ and $\delta {\tilde g}_{ab}$ are traceless. As discussed in the previous 
section, the functional derivatives 
with respect to ${\tilde A}^{ab}$ and ${\tilde g}_{ab}$ are trace free:
\begin{subequations} \label{gandAeqns}
\begin{eqnarray}
  \frac{\delta S^{(g)}}{\delta {\tilde g}_{ab}} & = & \frac{1}{2}\left( \dperp {\tilde A}^{ab} \right)^{\rm TF} 
  - \left( \frac{\delta H^{(g)}}{\delta{\tilde g}_{ab}} \right)^{\rm TF} \ ,\label{gandAeqns_g}\\
  \frac{\delta S^{(g)}}{\delta {\tilde A}^{ab}} & = & -\frac{1}{2}\left( \dperp {\tilde g}_{ab} \right)^{\rm TF} 
  - \left( \frac{\delta H^{(g)}}{\delta{\tilde A}^{ab}} \right)^{\rm TF} \ .\label{gandAeqns_A}
\end{eqnarray}
\end{subequations}
Here, $H^{(g)} \equiv \int d^3x \,\alpha {\cal H}^{(g)}$ is the gravitational field contribution to the Hamiltonian where 
${\cal H}^{(g)}$ is given by Eq.~(\ref{eqn:conformalHdensitytilde}).
The terms $\dperp {\tilde g}_{ab}$ and $\delta H^{(g)}/\delta{\tilde A}^{ab} = \alpha {\tilde A}_{ab}$ 
in Eq.~(\ref{gandAeqns_A}) 
are already traceless, so the equation of motion obtained by varying the action with respect to ${\tilde A}^{ab}$ is 
$\dperp {\tilde g}_{ab} = -2\alpha {\tilde A}_{ab}$. The term $\left( \dperp {\tilde A}^{ab} \right)^{\rm TF}$ can be written 
as 
  \begin{eqnarray} \label{dperpAcalculation}
    \left( \dperp {\tilde A}^{ab} \right)^{\rm TF} & = & \dperp {\tilde A}^{ab} - \frac{1}{3} {\tilde g}^{ab} 
    {\tilde g}_{cd} \dperp {\tilde A}^{cd} \nonumber\\
    & = & \dperp {\tilde A}^{ab} + \frac{1}{3} {\tilde g}^{ab} 
    {\tilde A}^{cd} \dperp {\tilde g}_{cd} \nonumber\\
    & = & \dperp {\tilde A}^{ab} - \frac{2}{3} \alpha {\tilde g}^{ab} 
    {\tilde A}^{cd} {\tilde A}_{cd} \ ,
  \end{eqnarray}
where the equation $\dperp {\tilde g}_{ab} = -2\alpha {\tilde A}_{ab}$ has been used. The evolution equation for 
${\tilde A}^{ab}$ is found by setting $\delta S^{(g)}/\delta{\tilde g}_{ab}$ equal to 
$-\delta S^{(m)}/\delta {\tilde g}_{ab} = -(\alpha {\tilde\varphi}^{16}/2) \left( s^{ab} \right)^{\rm TF}$ in 
Eq.~(\ref{gandAeqns_g}) and using the result from the calculation (\ref{dperpAcalculation}). The remaining 
equations of motion $\delta S/\delta{\tilde\varphi} = 0$ and $\delta S/\delta{\tilde\tau} = 0$ are 
straightforward to derive. The complete set is 
\begin{subequations}\label{invariantCTequations}
  \begin{eqnarray} 
    \dperp {\tilde g}_{ab}  & = & -2\alpha {\tilde A}_{ab} \ ,\label{caneoms_g}\\
    \dperp  {\tilde A}^{ab} & = &  2\alpha {\tilde A}^{ac}{\tilde A}_{c}^b 
       - 8{\tilde\varphi}^7 \left[ \alpha {\tilde D}^a {\tilde D}^b{\tilde\varphi}
       +  {\tilde D}^{(a}\alpha\, {\tilde D}^{b)}{\tilde\varphi} 
       + \frac{1}{8} {\tilde\varphi} {\tilde D}^a {\tilde D}^b\alpha 
       - \frac{1}{8} \alpha  {\tilde\varphi} {\tilde R}^{ab}
       + \frac{1}{8} \alpha  {\tilde\varphi}^{9} s^{ab} \right]^{\rm TF} \ ,\quad \label{caneoms_A}\\
    \dperp  {\tilde\varphi}  & = &  - \frac{1}{6} \alpha {\tilde\varphi}^7 {\tilde\tau} \ ,\label{caneoms_phi}\\
    \dperp {\tilde\tau}  & = &   \alpha  {\tilde\varphi}^6 {\tilde\tau}^2 
    +    \alpha {\tilde\varphi}^2 {\tilde R}
       - 14 \alpha {\tilde\varphi} {\tilde D}^2{\tilde\varphi} 
       -  {\tilde\varphi}^2 {\tilde D}^2\alpha 
       - 14 {\tilde\varphi} {\tilde D}^a{\tilde \varphi} {\tilde D}_a\alpha 
       - 42 \alpha  {\tilde D}^a{\tilde\varphi} {\tilde D}_a{\tilde\varphi} 
       + \frac{1}{2} \alpha {\tilde\varphi}^6 (s - 3\rho) \ .\quad\label{caneoms_tau}
  \end{eqnarray}
\end{subequations}
Note that the matter variables have not been conformally scaled, and the indices on the spatial stress tensor are 
lowered with the physical metric; in particular, $s = s^{ab} h_{ab}$.

The equations of motion for the original set of CT variables, $g_{ab}$, $A_{ab}$, $\varphi$, and $\tau$, 
can be found by extremizing the action (\ref{eqn:confaction1}). These equations are redundant, as implied 
by the relation in Eq.~(\ref{eomrelation}). 
Alternatively, we can obtain the independent equations of motion by inserting 
the definitions (\ref{canonicalCTvariables}) into Eqs.~(\ref{invariantCTequations}):
\begin{subequations} \label{CTequations}
  \begin{eqnarray}
    \dperp  g_{ab}  & = & \frac{1}{3} g_{ab}\, \dperp \ln g -2\alpha\sqrt{g}  A_{ab} \ ,\label{eoms_g}\\
    \dperp  A_{ab} & = & -\frac{1}{6}A_{ab}\, \dperp \ln g   -2\alpha \sqrt{g} A_{ac}A^c_{b} \nonumber\\ & &
        -8\sqrt{g} \varphi^7 \left[ \alpha  D_aD_b\varphi
       +  D_{(a}\alpha\,  D_{b)}\varphi + \frac{1}{8} \varphi D_aD_b\alpha - \frac{1}{8} \alpha  \varphi R_{ab}
       + \frac{1}{8} \alpha  \varphi s_{ab} \right]^{\rm TF} \ ,\quad \label{eoms_A}\\
    \dperp  \varphi & = & -\frac{1}{12} \varphi\, \dperp \ln g  
    - \frac{1}{6} \alpha \sqrt{g} \varphi^7 \tau \ ,\label{eoms_phi}\\
    \dperp  \tau  & = &   \sqrt{g} \Bigl[ \alpha  \varphi^6 \tau^2 + 
          \alpha  \varphi^2 R 
       - 14 \alpha  \varphi D^2\varphi -\varphi^2 D^2\alpha 
       - 14  \varphi D^a\varphi D_a\alpha 
       - 42 \alpha  D^a\varphi D_a\varphi + \frac{1}{2} \alpha \varphi^6 (s - 3\rho)\Bigr] 
    \ ,\quad\label{eoms_tau}
  \end{eqnarray}
\end{subequations}
These equations  are essentially identical to  Eqs.~(\ref{invariantCTequations}), apart from various factors of ${g}$ 
and the absence of tilde's. Note that the indices are up in Eq.~(\ref{caneoms_A}), 
and down in Eq.~(\ref{eoms_A}). This difference gives rise to the sign difference between the first term on the 
right--hand side of Eq.~(\ref{caneoms_A}) and the second term on the right--hand side of Eq.~(\ref{eoms_A}).

It might not be obvious that the terms involving the Ricci tensor ${\tilde R}_{ab}$ and covariant 
derivative ${\tilde D}_a$ on the right--hand sides of Eqs.~(\ref{caneoms_A}) and (\ref{caneoms_tau}) will 
simplify to the corresponding terms involving $R_{ab}$ and $D_a$ on the right--hand sides of Eqs.~(\ref{eoms_A}) 
and (\ref{eoms_tau}). However, the following argument shows that this must be the case. The terms involving 
${R}_{ab}$ and ${D}_a$ on the right--hand sides of Eqs.~(\ref{CTequations}) can be obtained from the
functional derivatives of
\begin{equation}
  F \equiv -\frac{1}{2} \int d^3 x \, \alpha h {\cal R} 
  = -\frac{1}{2} \int d^3 x \, \alpha g \left( \varphi^8 R - 8 \varphi^7 D^2 \varphi \right) \ .
\end{equation}
Since $F$ is conformally invariant, we can view it either as a functional of $g_{ab}$ and $\varphi$ or as a 
functional of the conformally invariant variables  $g^{1/3} g^{ab}$ and  $g^{1/12}\varphi$. The functional derivatives 
with respect to $g_{ab}$ and $\varphi$ are defined by 
\begin{equation}
  \delta F = \int d^3 x \left[ \frac{\delta F}{\delta g_{ab}} \delta g_{ab} + \frac{\delta F}{\delta \varphi} \delta\varphi 
  \right] \ .
\end{equation}
By splitting $\delta F/\delta g_{ab}$ into its trace and trace--free parts, we can rewrite this expression as
\begin{equation}
  \delta F = \int d^3 x \left[ - g^{-1/3} g_{ac} g_{bd} \left( \frac{\delta F}{\delta g_{cd}}\right)^{\rm TF} 
  \delta \left( g^{1/3} g^{ab}\right)  + g^{-1/12} \frac{\delta F}{\delta \varphi} \delta\left( g^{1/12} \varphi\right)  
  \right] \ .
\end{equation}
Then by explicit calculation, we find
\begin{subequations}\label{dF}
\begin{eqnarray}
  \frac{\delta F[g,\varphi]}{\delta (g^{1/3} g^{ab})} & = & -g^{-1/3} g_{ac} g_{bd} 
     \left( \frac{\delta F}{\delta g_{cd}} \right)^{\rm TF} \nonumber\\
       & = & -\frac{1}{2} g^{2/3} \varphi^7 \left[ \alpha \varphi R_{ab} - \varphi D_a D_b \alpha - 
             8 D_{(a}\alpha D_{b)} \varphi - 8 \alpha D_a D_b \varphi \right]^{\rm TF} \ , \label{dFdg}\\
  \frac{\delta F[g,\varphi]}{\delta (g^{1/12}\varphi)} & = & g^{-1/12} \frac{\delta F}{\delta \varphi} \nonumber\\
  & = & 4 g^{11/12} \varphi^5 \left[ -\alpha \varphi^2 R 
   + 14\alpha \varphi D^2\varphi
     + 42\alpha D^a\varphi D_a\varphi + 14\varphi D^a\varphi D_a\alpha + \varphi^2 D^2\alpha \right] \ .\label{dFdphi}
\end{eqnarray}
\end{subequations}
Since $F$,  $g^{1/3} g^{ab}$, and $g^{1/12}\varphi$ are conformally invariant, the expressions on 
the right--hand sides of Eqs.~(\ref{dFdg}) and (\ref{dFdphi}) must be conformally invariant. 
Therefore these expressions are unchanged if we 
conformally transform the CT variables to the invariant CT variables. We do this by applying the 
transformation (\ref{conformaltransformations}) with $\xi = g^{-1/12}$. This argument shows that we can 
place tilde's on the right--hand sides of Eqs.~(\ref{dF}) without changing the values of these expressions. 
Armed with this result, it is straightforward to show that Eqs.~(\ref{CTequations}) follow from 
Eqs.~(\ref{invariantCTequations}). 

The CT equations (\ref{CTequations}) extremize the action and are equivalent to the 
ADM equations (\ref{eqn:ADMequations}). They are invariant under the conformal transformation 
(\ref{conformaltransformations}). Thus, these equations do not determine the evolution of $g$, the determinant of 
the conformal metric $g_{ab}$. For practical (numerical) 
calculations, it can be useful to fix $g$. As discussed in the introduction, a common choice 
is to specify $g = 1$. 
However, we are free to choose $g$ to be any $t$--dependent spatial scalar density of weight $2$. There are two 
natural cases to consider for the evolution of $g$.  The first case is $\dperp g = 0$; 
that is, $g$ can be chosen to be constant along the normal to the spacelike hypersurfaces in spacetime. 
I will refer to this as the Eulerian condition, since $g$ is constant for the observers who are at 
rest in the spacelike hypersurfaces. The second case is $\partial g/\partial t = 0$; that is, $g$ is 
chosen to be constant along the time flow vector field in spacetime. I will
refer to this as the Lagrangian condition, since $g$ is constant for the observers who move along 
the ``flow lines'' defined by the spatial coordinates. 
Note that the choice $g = 1$ is a special case of the Lagrangian condition. 

For the Eulerian case the terms $\dperp \ln g$ vanish in the equations of motion (\ref{CTequations}).
Thus, we have 
\begin{subequations} \label{CTequations_Eulerian}
  \begin{eqnarray}
    \partial g_{ab}/\partial t  & = & 2D_{(a}\beta_{b)} -2\alpha\sqrt{g}  A_{ab} \ ,\label{eoms_g_E}\\
    \partial A_{ab}/\partial t  & = &  \beta^c D_c A_{ab} + 2 A_{c(a} D_{b)} \beta^c 
       -2\alpha \sqrt{g} A_{ac}A^c_{b} \nonumber\\
       & &  -8\sqrt{g} \varphi^7 \left[ \alpha  D_aD_b\varphi
       +  D_{(a}\alpha\,  D_{b)}\varphi + \frac{1}{8} \varphi D_aD_b\alpha - \frac{1}{8} \alpha  \varphi R_{ab}
       + \frac{1}{8} \alpha  \varphi s_{ab} \right]^{\rm TF} \ ,\quad \label{eoms_A_E}\\
    \partial\varphi/\partial t  & = &  \beta^c D_c\varphi - \frac{1}{6} \alpha \sqrt{g} \varphi^7 \tau \ ,\label{eoms_phi_E}\\
    \partial \tau/\partial t  & = &  \beta^c D_c\tau +  \sqrt{g} \Bigl[ \alpha  \varphi^6 \tau^2 + 
          \alpha  \varphi^2 R 
       - 14 \alpha  \varphi D^2\varphi - \varphi^2 D^2\alpha 
       - 14  \varphi D^a\varphi D_a\alpha 
       \nonumber\\  & & \qquad\qquad\qquad
       - 42 \alpha D^a\varphi D_a\varphi + \frac{1}{2} \alpha \varphi^6 (s - 3\rho) \Bigr] 
    \ ,\label{eoms_tau_E}
  \end{eqnarray}
\end{subequations}
when the condition $\dperp g = 0$ holds. 
For the Lagrangian case $\partial g /\partial t = 0$ we have
\begin{subequations} \label{CTequations_Lagrangian}
  \begin{eqnarray}
    \partial g_{ab}/\partial t  & = & \Bigl\{ {\hbox{right--hand side of Eq.~(\ref{eoms_g_E})}} \Bigr\} 
    - \frac{2}{3} g_{ab} D_c\beta^c   \ ,\label{eoms_g_L}\\
    \partial A_{ab}/\partial t  & = &  \Bigl\{ {\hbox{right--hand side of Eq.~(\ref{eoms_A_E})}} \Bigr\} 
      +  \frac{1}{3} A_{ab} D_c \beta^c   \ , \label{eoms_A_L}\\
    \partial\varphi/\partial t  & = &  \Bigl\{ {\hbox{right--hand side of Eq.~(\ref{eoms_phi_E})}} \Bigr\} 
     +  \frac{1}{6} \varphi D_c\beta^c   \ ,\label{eoms_phi_L}\\
    \partial \tau/\partial t  & = &   \Bigl\{ {\hbox{right--hand side of Eq.~(\ref{eoms_tau_E})}} \Bigr\} 
    \ .\label{eoms_tau_L}
  \end{eqnarray}
\end{subequations}
By contracting with $g^{ab}$ we find that Eq.~(\ref{eoms_g_E}) preserves the Eulerian condition. Likewise, 
Eq.~(\ref{eoms_g_L}) preserves the Lagrangian condition. 

The terms $2D_{(a} \beta_{b)}$ in Eqs.~(\ref{eoms_g_E}) and (\ref{eoms_g_L}) come from the Lie derivatives of the 
type ${0\choose 2}$ tensor $g_{ab}$.  The extra term 
$-(2/3)g_{ab}D_c\beta^c$ that appears in the Lagrangian 
case (\ref{eoms_g_L}) combines with $2D_{(a} \beta_{b)}$  to give the same result one 
would obtain by computing the Lie derivative as if $g_{ab}$ were a type ${0\choose 2}$ tensor density 
with weight $-2/3$. Similarly,  the extra terms that 
appear in the equations of motion for $A_{ab}$ and $\varphi$ in the Lagrangian case combine with the Lie derivatives 
${\cal L}_\beta A_{ab}$ and ${\cal L}_\beta \varphi$ to give the same results one would obtain by computing those 
Lie derivatives as if 
$A_{ab}$ were a type ${0\choose 2}$ tensor density of weight $1/3$ and $\varphi$ were a scalar density of weight $1/6$. 

As discussed in the introduction, there is no conceptual advantage in leaving the trace of $A_{ab}$ 
unspecified. If  $A=0$ were not imposed  from the beginning, the 
independent equations of motion would be identical 
to Eqs.~(\ref{CTequations}) with the replacements $\tau \to \tau + \varphi^{-6} A$ and 
$A_{ab} \to \left( A_{ab}\right)^{\rm TF}$. The resulting equations would not 
determine the evolution of $\tau$ and $A$ separately, 
but only the ``trace invariant'' combination $\tau + \varphi^{-6}A$. We could choose  $A$ to be any $t$--dependent 
spatial scalar. In practice it might be convenient to determine $A$ by removing the trace--free (TF) symbol from 
the terms in square brackets on the right--hand side of Eq.~(\ref{eoms_A}). 

\section{BSSN equations} 
The conformally invariant BSSN system of evolution equations \cite{Shibata:1995we,Baumgarte:1998te} 
is based on the CT equations (\ref{CTequations}). It uses  new fields, 
the ``conformal connection functions'', defined by
\begin{equation}\label{biggammadefinition}
  \Gamma^c \equiv g^{ab}\Gamma^c_{ab} = -\frac{1}{\sqrt{g}} \partial_a\left( \sqrt{g} g^{ac} \right) \ .
\end{equation}
Here, $\Gamma^c_{ab}$ are the Christoffel symbols built from the conformal metric $g_{ab}$ and
$\partial_a$ stands for the partial derivative with respect to the spatial coordinate $x^a$. 
In the original treatment of the BSSN system, the condition $g=1$ was imposed 
from the outset. As in the previous sections, I do not impose $g=1$. 

The key idea of the BSSN system is to replace certain derivatives 
of the metric that appear in the Ricci tensor 
on the right--hand side of Eq.~(\ref{eoms_A}) by the conformal connection functions $\Gamma^a$.  
By explicit calculation, we have 
\begin{equation}\label{ricci_bssn}
  R_{ab} = -\frac{1}{2} g^{cd} \partial_c \partial_d g_{ab} + g_{c(a}\partial_{b)}\Gamma^c + \Gamma^c \Gamma_{(ab)c}
      + g^{cd} \left( 2 \Gamma^e_{c(a} \Gamma_{b)ed} + \Gamma^e_{ac} \Gamma_{ebd} \right) \ ,
\end{equation}
where $\Gamma_{abc} \equiv g_{ad} \Gamma^d_{bc}$. The CT system (\ref{CTequations}) must be extended to include 
an evolution equation for $\Gamma^a$. This is found by computing  
$\dperp\Gamma^a$ from its definition (\ref{biggammadefinition}). 
For this calculation, it is helpful to introduce the invariant conformal connection functions, 
${\tilde\Gamma}^c \equiv {\tilde g}^{ab} {\tilde\Gamma}^c_{ab} = - \partial_a {\tilde g}^{ac}$.
These variables are invariant under the conformal transformation 
(\ref{conformaltransformations}) since they are built from ${\tilde g}_{ab} = g^{-1/3} g_{ab}$. They are 
related to $\Gamma^a$ by 
\begin{equation}\label{invariantbiggamma}
  {\tilde\Gamma}^a  = g^{1/3} \Gamma^a + (1/6)g^{-2/3} g^{ab} \partial_b g \ .
\end{equation}
The evolution equation for ${\tilde\Gamma}^a$ 
is straightforward to compute once we recognize that $\dperp$ and $\partial_a$ commute.\footnote{That is, 
${\cal L}_\beta$ and $\partial_a$ commute. To show this, recall that the components of, say, a contravariant 
vector in a ``barred'' 
coordinate system ${\bar x}^a$ are related to the components of that vector in an unbarred coordinate system by 
${\bar v}^a({\bar x}) = v^b(x) \partial_b {\bar x}^a$. The coordinate derivatives of the vector components are related by
${\bar\partial}_a {\bar v}^b({\bar x}) = ({\bar \partial}_a x^c) \partial_c [ v^d(x) \partial_d {\bar x}^b ]$. Similar 
expressions hold for the components, and derivatives of components, of other types of tensors. Define the Lie 
derivative by ${\cal L}_\beta T \equiv {\bar T}(x) - T(x)$ where ${\bar x}^a = x^a - \beta^a$ and $\beta$ is infinitesimal. 
(Indices have been suppressed on $T$.) One then finds that 
${\cal L}_\beta (\partial_a T) = \partial_a ({\cal L}_\beta T)$.} The result can be expressed  as
\begin{equation}\label{Gammaevolution}
  \dperp\left( g^{1/3} \Gamma^a \right) = -\frac{1}{6} \dperp \left( g^{1/3} g^{ab} \partial_b \ln g \right)
  -2  \partial_b \left( \alpha g^{5/6} A^{ab} \right) \ ,
\end{equation}
where $\partial_b \ln g = g^{-1} \partial_b g$. 

The equations (\ref{CTequations}), (\ref{Gammaevolution}) appear different from the 
familiar expression of the BSSN equations largely due to differences in notation. Let me define a set of ``BSSN variables'', 
denoted by carets:
\begin{subequations}\label{bssnvariables}
  \begin{eqnarray}
    \varphi & \equiv & e^{\hat \varphi} \ ,\\
    A_{ab} & \equiv & e^{6{\hat\varphi}} {\hat A}_{ab} \ ,\\
    \alpha & \equiv & e^{-6{\hat\varphi}} {\hat \alpha}/\sqrt{g} \ .
\end{eqnarray}
\end{subequations}
The definitions for the metric $g_{ab}$, trace of extrinsic curvature $\tau$, and shift vector $\beta^a$ are unchanged. 
Under a conformal transformation, the BSSN variables change by 
\begin{subequations}\label{bssnconformaltransformation}
  \begin{eqnarray}
    g_{ab} & \longrightarrow & \xi^4 g_{ab} \ ,\\
    {\hat \varphi} & \longrightarrow & {\hat\varphi} - \ln\xi \ ,\\
    {\hat A}_{ab} & \longrightarrow & \xi^4 {\hat A}_{ab} \ ,\\
    \tau & \longrightarrow & \tau \ ,\\
    \Gamma^a & \longrightarrow & \xi^{-4} \Gamma^a - 2 \xi^{-5} g^{ab} \partial_b\xi \ .
  \end{eqnarray}
\end{subequations}
The scalar lapse function ${\hat \alpha}$ and the shift vector $\beta^a$ are conformally invariant.

With the change of notation (\ref{bssnvariables}), the BSSN system (\ref{CTequations}), (\ref{Gammaevolution}) becomes 
\begin{subequations} \label{BSSNequations}
  \begin{eqnarray}
    \dperp  {g}_{ab}  & = & \frac{1}{3} g_{ab}\, \dperp\ln g -2{\hat\alpha} {\hat A}_{ab} \ ,\label{bssn_g}\\
    \dperp {\hat A}_{ab} & = & -\frac{1}{3} {\hat A}_{ab}\, \dperp \ln g 
        -2{\hat\alpha} {\hat A}_{ac}{\hat A}^c_{b} 
       + {\hat\alpha} {\hat A}_{ab} \tau   \nonumber\\
       & &  + e^{-4{\hat\varphi}}  \left[ -2{\hat\alpha} D_a D_b {\hat\varphi} 
	 + 4{\hat\alpha} D_a{\hat\varphi} D_b{\hat\varphi} + 4 D_{(a}{\hat\alpha} D_{b)} {\hat\varphi} 
	 - D_a D_b{\hat\alpha} + {\hat\alpha} R_{ab} - {\hat\alpha} s_{ab}  \right]^{\rm TF} \ ,\quad \label{bssn_A}\\
    \dperp  {\hat\varphi} 
        & = &  -\frac{1}{12} \, \dperp\ln g - \frac{1}{6} {\hat \alpha} {\tau} \ ,\label{bssn_phi}\\
    \dperp  \tau  & = &   {\hat \alpha} \tau^2
       + e^{-4{\hat\varphi}} \left( {\hat\alpha} R - 8 {\hat\alpha} D^2 {\hat\varphi} 
           - 8{\hat\alpha} D^a{\hat\varphi} D_a{\hat\varphi} - D^2{\hat\alpha} - 2 D^a{\hat\alpha} D_a{\hat\varphi} 
	   \right)  + \frac{1}{2}{\hat \alpha} (s - 3\rho)  \ ,\label{bssn_tau}\\
       \dperp \Gamma^a  & = & -\frac{1}{3} \Gamma^a\,\dperp\ln g 
        - \frac{1}{6} g^{ab} \partial_b \dperp\ln g  
       - \frac{2}{\sqrt{g}}  \partial_b \left( {\hat\alpha}\sqrt{g} {\hat A}^{ab} \right) \ .\label{bssn_Gamma}
  \end{eqnarray}
\end{subequations}
We can further modify these equations by making use of the Hamiltonian and momentum constraints 
${\cal H}^{(g)} = - g\varphi^{12} \rho$ and  ${\cal M}^{(g)}_a = \sqrt{g} \varphi^6 j_a$, 
where ${\cal H}^{(g)}$ and  ${\cal M}_a^{(g)}$ are given by Eqs.~(\ref{eqn:conformalHdensity}) 
and (\ref{conformalMdensity}), respectively. First, we rewrite the momentum constraint as 
\begin{equation}
  \partial_b \left( \sqrt{g}e^{6{\hat\varphi}} {\hat A}^{ab} \right) 
     = \sqrt{g} e^{6{\hat\varphi}} \left( -\Gamma^a_{bc} {\hat A}^{bc} 
     + \frac{2}{3} g^{ab}\partial_b\tau + g^{ab} j_b \right) \ ,
\end{equation}
and use this result to replace the spatial derivatives of ${\hat A}^{ab}$ on the right--hand side of Eq.~(\ref{bssn_Gamma}).
Next, we rewrite the Hamiltonian constraint as 
\begin{equation}
  R = e^{4{\hat\varphi}} \left( {\hat A}_{ab} {\hat A}^{ab} - \frac{2}{3} \tau^2 + 2\rho \right) 
    + 8\left(    D^2 {\hat\varphi} + D^a{\hat\varphi} D_a{\hat\varphi} \right) \ ,
\end{equation}
and use this result to replace the Ricci scalar $R$ on the right--hand side of Eq.~(\ref{bssn_tau}). The end result 
of these changes is 
\begin{subequations} \label{BSSNequations2}
  \begin{eqnarray}
    \dperp  {g}_{ab}  & = & \frac{1}{3} g_{ab}\, \dperp\ln g -2{\hat\alpha} {\hat A}_{ab} \ ,\label{bssn2_g}\\
    \dperp {\hat A}_{ab} & = & -\frac{1}{3} {\hat A}_{ab}\, \dperp \ln g 
        -2{\hat\alpha} {\hat A}_{ac}{\hat A}^c_{b} 
       + {\hat\alpha} {\hat A}_{ab} \tau   \nonumber\\
       & &  + e^{-4{\hat\varphi}}  \left[ -2{\hat\alpha} D_a D_b {\hat\varphi} 
	 + 4{\hat\alpha} D_a{\hat\varphi} D_b{\hat\varphi} + 4 D_{(a}{\hat\alpha} D_{b)} {\hat\varphi} 
	 - D_a D_b{\hat\alpha} + {\hat\alpha} R_{ab} - {\hat\alpha} s_{ab}  \right]^{\rm TF} \ ,\quad \label{bssn2_A}\\
    \dperp  {\hat\varphi} 
        & = &  -\frac{1}{12} \, \dperp\ln g - \frac{1}{6} {\hat \alpha} {\tau} \ ,\label{bssn2_phi}\\
    \dperp  \tau  & = &  \frac{1}{3} {\hat \alpha} \tau^2 + {\hat\alpha} {\hat A}_{ab} {\hat A}^{ab} 
        - e^{-4{\hat\varphi}} \left( D^2{\hat\alpha} + 2 D^a{\hat \alpha} D_a{\hat\varphi} 
	   \right) + \frac{1}{2}{\hat \alpha} (s + \rho)  \ ,\label{bssn2_tau}\\
       \dperp \Gamma^a  & = & -\frac{1}{3} \Gamma^a\,\dperp\ln g 
        - \frac{1}{6} g^{ab} \partial_b \dperp\ln g  - 2 {\hat A}^{ab} \partial_b{\hat\alpha} 
	+ 2{\hat\alpha} \left[ 6 {\hat A}^{ab}\partial_b{\hat\varphi} + \Gamma^a_{bc} {\hat A}^{bc} 
	  - \frac{2}{3} g^{ab}\partial_b\tau - g^{ab}j_b \right] 
      \ .\label{bssn2_Gamma}
  \end{eqnarray}
\end{subequations} 
Equations (\ref{BSSNequations2}) with the Ricci tensor $R_{ab}$ given by Eq.~(\ref{ricci_bssn}) define the 
conformally invariant BSSN system. 

We can rewrite the BSSN equations in a more compact form by using the identities
\begin{equation} \label{identity1}
  \nabla^2{\hat\alpha} = e^{-4{\hat\varphi}} \left( D^2{\hat\alpha} + 2 D^a{\hat\alpha} D_a{\hat\varphi} \right) 
\end{equation}
in Eq.~(\ref{bssn2_tau}) and  
\begin{equation} \label{identity2}
  \left[ {\hat\alpha} {\cal R}_{ab} - \nabla_a \nabla_b {\hat\alpha} \right]^{\rm TF} = 
  \left[  -2{\hat\alpha} D_a D_b {\hat\varphi} 
	 + 4{\hat\alpha} D_a{\hat\varphi} D_b{\hat\varphi} + 4 D_{(a}{\hat\alpha} D_{b)} {\hat\varphi} 
	 - D_a D_b{\hat\alpha} + {\hat\alpha} R_{ab} \right]^{\rm TF} 
\end{equation}
in Eq.~(\ref{bssn2_A}). Recall that ${\cal R}_{ab}$ and $\nabla_a$ are the Ricci tensor and covariant derivative 
constructed from the physical metric $h_{ab} = e^{4{\hat\varphi}} g_{ab}$. 

The BSSN equations (\ref{BSSNequations2}) are invariant under the conformal transformation 
(\ref{bssnconformaltransformation}).  As with the CT equations (\ref{CTequations}), these equations do
not determine the evolution of $g$. We must specify $\dperp g$, which appears in several places on the 
right--hand sides of Eqs.~(\ref{BSSNequations2}), as a separate condition. Alternatively, we could write 
the BSSN system in terms of conformally invariant variables, including the invariant conformal connection 
functions ${\tilde\Gamma}^a$.  This would remove the terms proportional to $\dperp \ln g$ in the BSSN equations but 
add certain density weights to each of the variables (except $\tau$). As defined here, $g_{ab}$ and ${\hat A}_{ab}$ 
are type $0\choose 2$ tensors with no density weight, and ${\hat \varphi}$ and $\tau$ are scalars with no 
density weight. The coordinate transformation rule for the conformal connection functions $\Gamma^a$ follow from 
their definition (\ref{biggammadefinition}) and the familiar inhomogeneous transformation rule for the 
Christoffel symbols. In particular, $\Gamma^a$ carries no density weight. 

Two natural choices for the evolution of $g$ are the Eulerian condition $\dperp g = 0$ and the Lagrangian 
condition $\partial g/\partial t = 0$. With the Eulerian condition, we have 
\begin{subequations} \label{BSSNequations_Euler}
  \begin{eqnarray}
    \partial  {g}_{ab}/\partial t  & = & 2 D_{(a}\beta_{b)} -2{\hat\alpha} {\hat A}_{ab} \ ,\label{bssn_g_E}\\
    \partial {\hat A}_{ab}/\partial t & = & \beta^c D_c{\hat A}_{ab} + 2{\hat A}_{c(a} D_{b)}\beta^c  
    -2{\hat\alpha} {\hat A}_{ac}{\hat A}^c_{b} 
       + {\hat\alpha} {\hat A}_{ab} \tau  + e^{-4{\hat\varphi}}  \left[ {\hat\alpha}{\cal R}_{ab} 
	 - \nabla_a\nabla_b{\hat \alpha} 
	 - {\hat\alpha} s_{ab}  \right]^{\rm TF} \ ,\quad \label{bssn_A_E}\\
    \partial  {\hat\varphi} /\partial t
        & = &  \beta^c D_c{\hat\varphi} - \frac{1}{6} {\hat \alpha} {\tau} \ ,\label{bssn_phi_E}\\
    \partial  \tau/\partial t  & = & \beta^c D_c\tau  +
    \frac{1}{3} {\hat \alpha} \tau^2 + {\hat\alpha} {\hat A}_{ab} {\hat A}^{ab} 
        - \nabla^2{\hat\alpha} + \frac{1}{2}{\hat \alpha} (s + \rho)  \ ,\label{bssn_tau_E}\\
       \partial \Gamma^a/\partial t  & = &  \beta^c \partial_c \Gamma^a - 
       \Gamma^c \partial_c \beta^a + g^{bc} \partial_b \partial_c \beta^a 
         - 2 {\hat A}^{ab} \partial_b{\hat\alpha} 
	+ 2{\hat\alpha} \left[ 6 {\hat A}^{ab}\partial_b{\hat\varphi} + \Gamma^a_{bc} {\hat A}^{bc} 
	  - \frac{2}{3} g^{ab}\partial_b\tau - g^{ab} j_b \right] 
      \ .\label{bssn_Gamma_E}
  \end{eqnarray}
\end{subequations}
The identities (\ref{identity1}) and (\ref{identity2}) have been used to express these equations in compact form. 
For the Lagrangian condition, we have
\begin{subequations} \label{BSSNequations_Lagrange}
  \begin{eqnarray}
    \partial  {g}_{ab}/\partial t  & = & \Bigl\{ {\hbox{right--hand side of Eq.~(\ref{bssn_g_E})}} \Bigr\} 
    - \frac{2}{3} g_{ab} D_c \beta^c 
    \ ,\label{bssn_g_L}\\
    \partial {\hat A}_{ab}/\partial t & = &  \Bigl\{ {\hbox{right--hand side of Eq.~(\ref{bssn_A_E})}} \Bigr\} 
    - \frac{2}{3} {\hat A}_{ab} D_c \beta^c \ , \label{bssn_A_L}\\
    \partial  {\hat\varphi} /\partial t
        & = &  \Bigl\{ {\hbox{right--hand side of Eq.~(\ref{bssn_phi_E})}} \Bigr\} + \frac{1}{6} D_c \beta^c
    \ ,\label{bssn_phi_L}\\
    \partial  \tau/\partial t  & = &  \Bigl\{ {\hbox{right--hand side of Eq.~(\ref{bssn_tau_E})}} \Bigr\} 
    \ ,\label{bssn_tau_L}\\
       \partial \Gamma^a/\partial t  & = &   \Bigl\{ {\hbox{right--hand side of Eq.~(\ref{bssn_Gamma_E})}} \Bigr\} 
       + \frac{2}{3} \Gamma^a D_c \beta^c + \frac{1}{3} D^a D_c \beta^c
      \ .\label{bssn_Gamma_L}
  \end{eqnarray}
\end{subequations}
The Eulerian and Lagrangian conditions are preserved by equations (\ref{bssn_g_E}) and (\ref{bssn_g_L}), respectively. 
Note that the restriction $g=1$ is a special case of the Lagrangian condition. Also note that with $g=1$, the 
extra terms in Eqs.~(\ref{BSSNequations_Lagrange}) simplify since in that case $D_c\beta^c = \partial_c \beta^c$. 

It has been suggested \cite{Baumgarte:1998te} that the ``Gamma freezing''  condition, 
$\partial\Gamma^a/\partial t = 0$ might be useful as a means of specifying the 
shift vector for numerical evolutions  based on the BSSN system. The related ``Gamma driver'' conditions 
\cite{Alcubierre:2002kk}, which are 
less time--consuming to solve numerically than the Gamma freezing condition,  have been used with some success. Here 
we note that 
these conditions depend specifically on the way that the shift terms enter the evolution equation for $\Gamma^a$. If we 
choose the Eulerian condition $\dperp g = 0$ to break the conformal invariance,  then the Gamma freezing shift equation is 
\begin{equation} 
  g^{bc} \partial_b\partial_c \beta^a = {\hbox{terms containing at most first derivatives of $\beta^a$}} \ .
\end{equation}
If we choose the Lagrangian condition $\partial g /\partial t = 0$ to break conformal invariance, the Gamma 
freezing condition becomes 
\begin{equation}
  g^{bc}\partial_b\partial_c \beta^a + \frac{1}{3} g^{ab} \partial_b \partial_c \beta^c = 
  {\hbox{terms containing at most first derivatives of $\beta^a$}} \ .
\end{equation}
It might be interesting to explore the difference between these two shift conditions.

\section{Conformal thin sandwich  equations}
Initial data for general relativity must satisfy the constraint equations. In the original York--Lichnerowicz conformal 
decomposition \cite{Cook:2000vr}, the gravitational field parts of the constraints are written as in 
Eqs.~(\ref{eqn:conformalHdensity}) and
(\ref{conformalMdensity}). The Hamiltonian constraint ${\cal H}^{(g)} = - \varphi^{12} g \rho$ can be solved for the 
Laplacian of the conformal factor $\varphi$, 
\begin{equation}\label{YorkHconstraint}
  D^2 \varphi = -\frac{1}{8} \varphi^{-7} A^{ab}A_{ab} + \frac{1}{12} \varphi^5 \tau^2 
  + \frac{1}{8} \varphi R - \frac{1}{4}\varphi^5 \rho \ .
\end{equation}
The trace free extrinsic curvature $A_{ab}$ is split into a transverse part and a longitudinal 
part. With the so--called ``conformal transverse--traceless decomposition'', the longitudinal part of $A_{ab}$ 
is expressed in terms of derivatives of a vector $X^a$. Then the principal 
part of the momentum constraint ${\cal M}_a = \varphi^6 \sqrt{g} j_a$ is proportional to 
\begin{equation}
  \Delta_\LL X_a  \equiv  D_b (\LL X)^{b}_a \ ,  \label{yorkoperator_a}
\end{equation}
where
\begin{equation}
   (\LL X)_{ab}  \equiv   2 D_{(a} X_{b)} - (2/3) g_{ab} D_c X^c  \ .\label{yorkoperator_b}
\end{equation}
$\Delta_\LL$ is an elliptic 
operator. In this way, the Hamiltonian and momentum constraints are expressed as a system of elliptic 
equations for $\varphi$ and $X^a$. The freely specifiable parts of the gravitational field are the conformal metric 
$g_{ab}$, the trace of the extrinsic curvature $\tau$, and the transverse part of $A_{ab}$. 

More recently, York \cite{York:1998hy} has recognized that the momentum constraint can be expressed in terms of the elliptic 
operator $\Delta_\LL$ acting on the shift vector. 
Observe that the trace--free part of the conformal metric 
velocity is $\left( \partial g_{ab}/\partial t \right)^{\rm TF} 
= \partial g_{ab} / \partial t - (g_{ab}/3)\, \partial(\ln g)/\partial t$. 
Using the ``dot'' notation for time derivatives, we find that
Eq.~(\ref{eoms_g}) becomes 
\begin{equation} \label{newAeqn}
  A_{ab} = -\frac{1}{2\alpha \sqrt{g}} \left( {\dot g}_{ab}^{\rm TF} - (\LL \beta)_{ab} \right) 
\end{equation}
where $(\LL \beta)_{ab}$ is defined in Eq.~(\ref{yorkoperator_b}). 
Inserting this result into the momentum constraint, we find 
\begin{equation} \label{YorkMconstraint}
 \Delta_\LL \beta_a  = D^b {\dot g}^{\rm TF}_{ab} + 2 \sqrt{g} A_{ab} D^b\alpha 
  + \frac{4}{3} \alpha \sqrt{g} \varphi^6 D_a\tau + 2 \alpha \sqrt{g} \varphi^6 j_a \ ,
\end{equation}
where  $\Delta_\LL$ is defined by Eq.~(\ref{yorkoperator_a}).
As described in Ref.~\cite{York:1998hy}, one can specify freely the gravitational quantities $g_{ab}$, 
${\dot g}^{\rm TF}_{ab}$, $\tau$, and $\alpha$ then solve Eqs.~(\ref{YorkHconstraint}) and (\ref{YorkMconstraint}) 
for $\varphi$ and $\beta^a$. Wherever $A^{ab}$ appears, it is written in terms of the conformal metric 
and its derivatives via Eq.~(\ref{newAeqn}). 

The conformal thin sandwich construction \cite{Pfeiffer:2002iy} is an extension of 
this analysis to include the lapse anti--density as one of the unknowns. 
By using the Hamiltonian constraint (\ref{YorkHconstraint}) to eliminate the Laplacian of 
$\varphi$, we find that Eq.~(\ref{eoms_tau}) can be written as 
\begin{eqnarray}\label{alphaeqn}
  D^2\alpha & = & -\frac{1}{\sqrt{g}} \varphi^{-2} \dperp\tau - \frac{1}{6} \alpha \varphi^4 \tau^2
    - \frac{3}{4} \alpha R + \frac{7}{4} \alpha \varphi^{-8} A^{ab}A_{ab}  \nonumber\\
    & & - 14\varphi^{-1} D^a\varphi D_a\alpha 
   - 42 \alpha \varphi^{-2} D^a\varphi D_a\varphi + \frac{1}{2} \alpha \varphi^4 (s + 4\rho) \ .
\end{eqnarray}
The principal part of this equation is the conformal Laplacian operator acting on $\alpha$. 
For the conformal thin sandwich initial data construction the freely specified quantities are the 
conformal metric $g_{ab}$, the trace--free part of the metric velocity 
${\dot g}_{ab}^{\rm TF}$, the trace of the 
extrinsic curvature $\tau$, and its time derivative ${\dot \tau} = \partial \tau /\partial t$.  
Equations (\ref{YorkHconstraint}),  (\ref{YorkMconstraint}), and (\ref{alphaeqn}), along with the 
expression for $A^{ab}$ given in Eq.~(\ref{newAeqn}), constitute an 
elliptic system of equations to be solved 
for the conformal factor $\varphi$, the shift vector $\beta^a$, and the lapse anti--density $\alpha$. 
Once these equations are solved, the initial data in terms of the physical metric $h_{ab}$ and
extrinsic curvature $K_{ab}$ can be obtained from Eqs.~(\ref{ctdecomposition}). 

The conformal thin sandwich equations (\ref{YorkHconstraint}), (\ref{newAeqn}), (\ref{YorkMconstraint}),
and (\ref{alphaeqn}) 
are derived from the CT equations  and the Hamiltonian and momentum constraints, all of which are 
conformally invariant. It follows that the conformal thin sandwich equations are invariant under the 
conformal transformation (\ref{conformaltransformations}). To be precise, let $\varphi$, $\beta^a$, and 
$\alpha$ denote the 
solution of the conformal thin sandwich equations for a given set of input data $g_{ab}$, 
${\dot g}^{\rm TF}_{ab}$, $\tau$, and 
${\dot\tau}$. Then the solution for input data $\xi^4 g_{ab}$, $\xi^4 {\dot g}^{\rm TF}_{ab}$, $\tau$, and 
${\dot\tau}$ will be $\varphi/\xi$, $\beta^a$, and $\alpha$. The physical metric $h_{ab}$ and 
extrinsic curvature $K_{ab}$ are the same in either case.

In evolving the conformal thin sandwich data one can choose
the lapse anti--density and shift vector $\alpha$ and $\beta^a$ freely, without regard to the 
values obtained from  the initial data construction. However, if  the initial values of the lapse 
and shift are chosen to coincide with the values of $\alpha$ and $\beta^a$ that were 
computed from the thin sandwich equations,  then initially the  trace--free part of the 
conformal metric velocity will coincide with the chosen value of ${\dot g}^{\rm TF}_{ab}$ and the 
initial time derivative of the  extrinsic curvature's trace will be given by the chosen value 
of ${\dot\tau}$. This will be the case whether the data is evolved via the ADM equations 
(\ref{eqn:ADMequations}), the CT equations (\ref{CTequations}), or the BSSN equations 
(\ref{BSSNequations2}). In the later two cases, we are free to 
choose the Eulerian or Lagrangian condition, or any other condition for the evolution of $g$.  

\section{CFC equations}
The approximate evolution equations obtained from the CFC have been 
used in numerical studies of  binary neutron star coalescence \cite{Wilson:1996ty,Marronetti:1998xv} and supernovae 
\cite{Dimmelmeier:2002bk,Dimmelmeier:2004me}. This approximation to general relativity appears to be quite 
good \cite{Cook:1995cp}, at least for systems that are not too far from spherical symmetry. 

It turns out that the CFC equations are precisely the  conformal thin sandwich equations, 
(\ref{YorkHconstraint}), (\ref{newAeqn}), (\ref{YorkMconstraint}), and (\ref{alphaeqn}). The CFC 
approximation can be described in a generalized way as follows. Let the conformal metric $g_{ab}$ and 
the trace of the extrinsic curvature $\tau$ be specified freely for all times. Then $g_{ab}$, 
${\dot g}_{ab}^{\rm TF}$, 
$\tau$, and ${\dot \tau}$ are known for all times and the thin sandwich equations can be solved 
for the gravitational data $\alpha$, $\beta^a$, $\varphi$, and $A_{ab}$ as soon as the source 
data $\rho$, $j_a$, and $s^{ab}$ is available. The idea, then, is to solve the conformal 
thin sandwich equations at the initial time using the initial values for the sources. We then evolve the sources 
forward in time to the first timestep using 
the matter equations of motion. These equations depend on the physical metric $h_{ab} = \varphi^4 g_{ab}$, lapse 
anti--density, and shift. From the source values at the first timestep, the conformal thin sandwich equations 
are solved to complete the gravitational data at the first timestep. The process repeats. 

For the CFC approximation, one sets the conformal metric equal to a flat metric for all times. In particular 
${\dot g}_{ab}^{\rm TF} = 0$ in Eq.~(\ref{YorkMconstraint}). Also, maximal slicing is assumed so that 
$\tau = 0$ for all times. Recall that the conformal thin sandwich equations are equivalent to the 
Hamiltonian constraint, the momentum constraint, and two of the four CT evolution equations, namely, 
Eqs.~(\ref{eoms_g}) and (\ref{eoms_tau}). Thus, the approximation in the CFC formalism consists in 
ignoring the evolution 
equations for  $A_{ab}$ and $\varphi$, namely, Eqs.~(\ref{eoms_A}) and (\ref{eoms_phi}). 

\section{Summary}
The conformal--traceless decomposition of the gravitational field appears in a number of contexts in general relativity,
most notably in the analysis of the initial value 
problem and in the construction of the BSSN system of evolution equations. In this paper I have written 
the action functional and  equations of motion in terms of conformal--traceless variables. I do not invoke the 
frequently imposed 
condition $g=1$ on the determinant of the conformal metric. As a consequence, the action and equations of 
motion are conformally invariant. I have presented two possibilities for breaking conformal invariance, namely, the 
Eulerian condition $\dperp g = 0$ and Lagrangian condition $\partial g/\partial t = 0$ on the evolution of $g$. 
I also extended the equations of motion to obtain 
a conformally invariant version of the  BSSN system. For this system as well, the invariance can be 
broken by specifying the Eulerian or Lagrangian condition. I showed that the conformal thin sandwich equations 
for gravitational initial data
are obtained from the Hamiltonian and momentum constraints along with a subset of the evolution equations 
written in conformal--traceless variables. Finally, I have pointed out that the CFC approximation to general relativity 
consists in solving the conformal thin sandwich equations at each time step, assuming the conformal metric is 
flat and the time slicing is maximal.

\begin{acknowledgments}
I would like to thank to J.W.~York for countless discussions through the years. This work was supported by 
NASA Space Sciences Grant ATP02-0043-0056 and by NSF Grant PHY-0070892. 
\end{acknowledgments}

\bibliography{references}
\end{document}